\documentclass[preprint,aps,amssymb,eqsecnum,nofootinbib]{revtex4}
\usepackage{graphicx}
\usepackage{amsmath}
\usepackage{amssymb}
\usepackage{bm}
\usepackage{color}
\usepackage{verbatim}
\usepackage{feynmp}
\usepackage{mathrsfs}

\def\be{\begin{equation}}
\def\ee{\end{equation}}
\def\ba{\begin{eqnarray}}
\def\ea{\end{eqnarray}}

\def\mpl{M_{\rm p}}
\def\be{\begin{equation}}
\def\ee{\end{equation}}
\def\bi{\begin{itemize}}
\def\ei{\end{itemize}}
\def\ben{\begin{enumerate}}
\def\een{\end{enumerate}}
\def\bt{\begin{tabular}}
\def\et{\end{tabular}}
\def\bc{\begin{center}}
\def\ec{\end{center}}
\def\la{\label}

\def\bea{\begin{eqnarray}}
\def\eea{\end{eqnarray}}
\def\l{\left}
\def\r{\right}
\def\f{\frac}

\def\d{\partial}

\def\fr{\frac}
\def\la{\label}

\newcommand{\mbf}[1]{\mathbf{#1}}
\def\vphi{\varphi}
\def\pr{\prime}

\begin{document}

\title{Where does Cosmological Perturbation Theory Break Down?}
\author{Cristian Armendariz-Picon}
\author{Michele Fontanini}
\author{Riccardo Penco}
\author{Mark Trodden}

\affiliation{Department of Physics, Syracuse University, Syracuse, NY 13244, USA}

\begin{abstract}
We apply the effective field theory approach to the coupled metric-inflaton system, in order to investigate the impact of higher dimension operators on the spectrum of scalar and tensor perturbations in the short-wavelength regime. In both cases, effective corrections at tree-level become important when the Hubble parameter is of the order of the Planck mass, or when the physical wave number of a cosmological perturbation mode approaches the square of the Planck mass divided by the Hubble constant. Thus, the cut-off length below which conventional cosmological perturbation theory does not apply is likely to be much smaller than the Planck length. This has implications for the observability of ``trans-Planckian" effects in the spectrum of primordial perturbations.
\end{abstract}
\maketitle

\section{Introduction} \la{intro}

\begin{fmffile}{graphs}

One of the main successes of inflation~\cite{inflation} is the explanation of the origin of structure~\cite{structure}. During slow-roll, the Hubble radius remains nearly constant, while cosmological modes  are constantly pushed out of the horizon. Thus, local processes determine the amplitude and properties of perturbations at sub-horizon scales, which are transfered to  cosmologically large distances by the accelerated expansion.  In that sense, the sky is the screen upon which inflation has projected the physics of the microscopic universe.

The primordial perturbations seeded during inflation arise from quantum-mechanical fluctuations of the inflaton around its homogeneous value. Hence, their properties directly  depend on the quantum state of the inflaton perturbations.  Conventionally, this is taken to be  a state devoid of quanta in the asymptotic past, raising the crucial question of whether we can trust cosmological perturbation theory---and its quantum nature---at such early times~\cite{Brandenberger:1999sw}.

According to our present understanding, quantum field theories and general relativity are merely low energy descriptions of a more fundamental theory of quantum gravity. In the case of inflation, the leading terms in the corresponding effective Lagrangian are the Einstein-Hilbert term plus the inflaton kinetic term and potential. In an effective field theory treatment, these terms are accompanied by all other possible operators compatible with the symmetries of the theory, namely, general covariance and any other symmetry of the inflaton sector.  Higher dimensional operators are suppressed by powers of an energy scale,  which we will assume to be of the order of the reduced Planck mass $\mpl \equiv (8 \pi G)^{-1/2}$, and they are therefore expected to be negligible at sufficiently small momenta, or sufficiently long wavelengths.  The goal of this paper is to determine the three-momentum scale $\Lambda$ at which  these operators become important. The terms that yield the  leading (momentum-independent) corrections to  the primordial spectrum have been recently discussed  in \cite{Weinberg:2008hq}.  

It is crucial to realize that $\Lambda$ does not need to equal the Planck scale. As an extreme example, let us consider perturbations around a background given by a constant scalar field in Minkowski spacetime, $\vphi = \varphi_0+\delta \vphi$ and $g_{\mu \nu} = \eta_{\mu \nu} + \delta g_{\mu \nu}$.
In this case, the effective action for the perturbations  is Lorentz invariant, because the background is Lorentz invariant and we assume that the original effective action $S[\varphi,g_{\mu\nu}]$ is generally covariant (we are implicitly assuming that a Lorentz-invariant gauge on $\delta g_{\mu\nu}$ has been imposed.) Therefore, the regime in which higher order operators become important will be expressed by  Lorentz invariant relations such as, for example, $k_{\mu} k^{\mu} \approx \mpl^2$. In other words, there is not any bound on the three-momentum $|\mathbf{k}|$, simply because a relation like $|\mathbf{k}| \approx \mpl$ is not Lorentz invariant.   One would like to extend this argument to an expanding universe. On short time-scales and short distances, spacetime can be regarded as flat and the scalar field driving inflation can be regarded as constant. Hence, it seems that cosmological perturbation theory should be valid all the way into the regime $|\mathbf{k}|\to \infty$, because the relevant four-momenta of cosmological perturbations are light-like, $k_\mu k^\mu=0$. As we shall see though, the evolution of the inflaton leads to small but finite violations of the Lorentz symmetry even in the short-wavelength limit,  which are imprinted on the power spectrum as $\mathbf{k}$-dependent corrections. 

The phenomenological imprints  of trans-Planckian physics on the primordial spectrum of perturbations, and the  implications of a finite cut-off $\Lambda$ on the spatial momentum of cosmological modes have been extensively studied  \cite{Martin:2000xs,Brandenberger:2000wr,Niemeyer:2000eh,Kempf:2000ac,Starobinsky:2001kn,Easther:2001fi,Easther:2001fz,Niemeyer:2001qe,Kempf:2001fa,Brandenberger:2002hs,Danielsson:2002kx,Easther:2002xe,Bergstrom:2002yd,Kaloper:2002uj,ArmendarizPicon:2003gd,Martin:2003kp,Elgaroy:2003gq,Martin:2003sg,Okamoto:2003wk,Shankaranarayanan:2004iq,Martin:2004yi,Schalm:2004qk,Greene:2004np,Schalm:2004xg,Easther:2004vq,Greene:2005aj}.  These articles mostly study corrections to the power spectrum in  the long-wavelength limit $k/a \equiv k_{ph}\ll H$, at late times, which is the regime directly accessible by experimental probes. In this article we focus instead on the short-wavelength regime $k_{ph}\gg H$, at early times, since we are interested in determining how far into the  ultraviolet cosmological perturbation theory applies. At short wavelengths, the power spectrum can be cast as a derivative expansion of the form
\begin{equation}
\langle \delta\varphi^*(\mathbf{k}) \delta\varphi(\mathbf{k})\rangle=\frac{1}{2k}\left(1+\alpha_2\frac{k_{ph}^2}{ \mpl^2}+\alpha_4\frac{k_{ph}^4}{\mpl^4}+\cdots\right),
\end{equation}
with coefficients $\alpha_i$ that depend on slow-roll parameters and the dimensionless ratio $H/\mpl$.    The analytic corrections to the leading result $1/2k$ arise from tree-level diagrams with vertices  from higher-dimensional operators. 
We only consider tree-level diagrams here, since we expect loop diagrams to simply introduce a logarithmic dependence of the dimensionless coefficients $\alpha_i$ on scale. Cosmological perturbation theory fails when the expansion in powers of $k_{ph}$ breaks down, namely, when all the terms become of the same order,
\begin{equation}
	k_{ph} \approx \mpl \sqrt{\f{\alpha_{2n}}{\alpha_{2n + 2}}} \equiv \Lambda \ .
\end{equation}
As we shall show,  the ratios $\alpha_{2n}/\alpha_{2n+2}$ are all quite large and of the same order, so the effective cut-off $\Lambda$ significantly differs from $\mpl$. In a slightly different context, a similar analysis  has been applied to the bispectrum in \cite{Holman:2007na}.

The structure of this paper is as follows. In the next section we describe the relevant background to our problem, setting up a description of perturbation theory in cosmology and describing the basics of tensor and scalar fluctuations. In section \ref{sec:tensors} we compute the squared amplitude of tensor perturbations and we derive the results mentioned above. In section~\ref{sec:scalars} we apply a similar analysis to the case of scalar perturbations, and obtain similar results. We conclude and discuss possible implications of our results in section~\ref{sec:conclusions}. Throughout this paper we work in units such that $\hbar \! = \! c \! = \!1$, and with metric signature $(-+++)$.

\section{Setting the scene -- Cosmological Perturbation Theory}
\label{sec:background}

\subsection{The Inflating Background}
Our starting point is a standard single-field inflation model. At sufficiently late times, the inflaton and gravity must be described by a low-energy effective action, whose leading terms are dictated by general covariance and the field content,
\be \la{1}
S_0 = \int \,d^4 x\sqrt{-g} \l[ \f{\mpl^2}{2} R - \frac{1}{2} \d_{\mu} \vphi \, \d^{\mu} \vphi - V(\vphi) \r] .
\ee
In an effective field theory context, the action should also contain additional terms suppressed by powers of a dimensionful scale, which we assume to be of the order of the Planck mass $\mpl$. Our goal is to determine the point beyond which such higher-dimensional operators produce corrections to the two-point function of cosmological perturbations that cannot be neglected. Our considerations can be readily generalized to cases in which the suppression scale of the higher-dimensional operators is not the Planck mass, but any other scale. 

If the potential $V(\vphi)$ is sufficiently flat, at  least in a certain region in field space, there exist inflationary solutions, along which a homogeneous scalar field $\vphi (\eta, \mbf{x})= \vphi_0 (\eta)$ slowly rolls down the potential and spacetime is  spatially homogeneous, isotropic and flat,
\be
g_{\mu \nu}^{(0)} \equiv a^2(\eta) \eta_{\mu\nu}, 
\ee
where $\eta_{\mu\nu}$ is the Minkowski metric and $\eta$ denotes conformal time.  A model-independent measure of the slowness of the inflation is given by the slow-roll parameter
\begin{equation}
	\epsilon\equiv - \f{H'}{a H^2} , 
\end{equation}
where $H \equiv a'/a^2$ is the Hubble parameter and a prime denotes a derivative with respect to conformal time. During slow-roll, $\epsilon$ is nearly constant, and to lowest order in slow-roll parameters its time derivative can be neglected.  Throughout this article we work to  leading non-vanishing order in the slow-roll expansion.

\subsection{Cosmological  Perturbations}

Let us now consider cosmological perturbations around the homogeneous and isotropic background described above. Writing $\varphi=\varphi_0+\delta\varphi$ and $g_{\mu\nu}=g_{\mu \nu}^{(0)}(\eta) +\delta g_{\mu\nu}(\eta, \mbf{x})$, and substituting into equation~(\ref{1}), we can expand the action $S_0$ up to the desired order in the fluctuations $\delta\varphi$ and $\delta g_{\mu\nu}$,
\begin{equation}
S_0[\varphi, g_{\mu\nu}] = \delta_0 S_0 + \delta_1 S_0 + \delta_2 S_0 + \cdots .
\end{equation}
The lowest order term $\delta_0 S_0$ does not contain any fluctuations and describes the inflating background; the linear term $\delta_1 S_0$ vanishes because it corresponds to the first variation of the action along the background solution, and the quadratic part of the action  $\delta_2 S_0$ describes the free dynamics of the perturbations. The latter is what we need  in order to  calculate the primordial spectrum of fluctuations.  To quadratic order, tensor and scalar perturbations are decoupled, so we may study them separately.

\subsubsection{Tensor Perturbations}

Tensor perturbations are described by a transverse and traceless tensor $h_{ij}$,
\begin{equation}
ds^2 = a^2(\eta) \l[ -d\eta^2 + (\delta_{i j} + h_{i j}) dx^i dx^j \r]. \la{3}
\end{equation}
The tensor $h_{ij}$ itself can be decomposed in plane waves of two different polarizations, which again decouple at quadratic order. We shall hence focus on just one of them,
\be\label{eq:tensor expansion}
h_{i j} (\eta, \mbf{x}) = \frac{1}{\sqrt{V}} \sum_{\mathbf{k}}  e_{i j} (\mbf{k}) \, h_k (\eta) \, e^{i \mbf{k} \cdot \mbf{x}},
\ee
where the $h_k (\eta)$ are the corresponding mode functions, and $e_{ij}(\mbf{k})$ denotes the normalized graviton polarization tensor,  $e^i{}_j e^j{}_i=1$ (we raise and lower spatial indices with the Kronecker delta.) Note that for later convenience we work in a toroidal universe of volume $V=L^3$; hence, the spatial wave numbers   have components $k_i=n_i \cdot (2\pi/L)$, where the $n_i$ are arbitrary integers.

Substituting the expansion (\ref{eq:tensor expansion}) into the action (\ref{1}), and using the background equations of motion, we may then express the free action $\delta_2 S_0$ as 
\be \la{5}
\delta_2 S_0 = \f{1}{2} \int d \eta \,\sum_{\mathbf{k}}  \l[ v^{\pr}_k v^{\pr}_{- k} -\l( k^2 - \f{a^{\pr \pr}}{a} \r) v_k v_{- k} \r] \ ,
\ee
where the scalar variable $v_k$ is defined as
\be \la{11}
v_k = a\, \mpl  \, h_k \ .
\ee
Thus, in terms of $v_k$ the action for tensor perturbations takes the form of an harmonic oscillator with time-dependent frequency. 

\subsubsection{Scalar Perturbations}

In spatially flat gauge, the perturbed metric reads
\be
ds^2 = a^2(\eta) \l[ -(1 + 2 \phi) d\eta^2 + 2 \d_i B dx^i d\eta + \delta_{i j} dx^i dx^j \r] \ . \la{metricpert}
\ee
On first inspection there appear to be three independent scalar variables: $\phi$, $B$ and the inflaton perturbation $\delta\varphi$. However, Einstein equations impose constraints on both $\phi$ and $B$. Solving the corresponding Fourier transformed equations to leading order in the slow-roll expansion, one finds (see e.g. \cite{Hu:2004xd})
\begin{subequations}\label{eq:constraints}
\bea
\phi_k &=& \sqrt{\fr{\epsilon}{2}} \fr{\delta\varphi_k}{\mpl} \ , \la{60}\\
B_k &=& \sqrt{\fr{\epsilon}{2}} \fr{\delta\vphi_k^\pr}{\mpl k^2} \ .  \la{61}
\eea
\end{subequations}
Consequently,  there is only one physical scalar degree of freedom, and scalar perturbations can be described by just one variable. A convenient choice that is particularly useful for quantizing scalar perturbations is the Mukhanov variable~\cite{Mukhanov:1988jd}, which in spatially flat gauge takes the simple form
\be \la{38}
v_k = a \, \delta\vphi_k \ .
\ee
Using relations (\ref{eq:constraints}) and (\ref{38}), we may express $\delta_2 S_0$ in terms of $v_k$ only. For constant $\epsilon$, that is, to leading order in the slow-roll expansion, the resulting action is  also given by equation (\ref{5}). This agreement greatly simplifies the analysis, because it allows us to use the same set of propagators to describe both scalar and tensor fluctuations. 

To leading order in the slow-roll expansion, the mode functions of both scalar and tensor perturbations hence satisfy the same equation of motion during inflation. Varying the action (\ref{5}) with respect to $v_{-k}$ we obtain
\be\la{30}
v_k^{\pr \pr} + \l[ k^2 -\frac{a''}{a}\r] v_k = 0 \ ,
\ee
which has a unique solution for appropriate initial conditions. The conventional choice is the Bunch-Davies or adiabatic vacuum, whose mode functions obey
\be \la{20}
v_k (\eta) \quad \stackrel{| k \eta| \gg 1}{\longrightarrow} \quad \f{e^{- i k \eta}}{\sqrt{2k}}\l[1+\mathcal{O}\l(\frac{1}{k\eta}\r)\r] \ .
\ee
Because we are only interested in the sub-horizon limit, this is all we need to know about the mode functions. In particular, because the behavior of the mode functions in the short-wavelength limit does not depend on the details of inflation, our results are also insensitive to the particular form of the inflaton potential. 

\subsection{Quantum Fluctuations and the {\it in-in} Formalism}
\la{inin}
In order to study the properties of cosmological modes in the short-wavelength regime, we concentrate on the  two-point function of the field $v$,
\be \la{50}
\langle v^* (\eta, \mbf{k}) v (\eta, \mbf{k}) \rangle \equiv \langle 0, in | v^* (\eta, \mbf{k}) v (\eta,  \mbf{k}) | 0, in \rangle \ ,
\ee
where $|0,in\rangle$ is the quantum state of the perturbations, which we assume to be the Bunch-Davies vacuum. The two-point function characterizes the mean square amplitude of cosmological perturbation modes, and differs from  the power spectrum just by a normalization factor. Note that in an infinite universe, the two-point function is proportional to a momentum-conserving delta function, which in a spatially compact universe is replaced by a Kronecker delta.

In the {\it in-in} formalism (see \cite{inin} for a clear and detailed exposition) the two-point function can be expressed as a path integral, 
\ba\label{eq:path integral}
&& \!\!\!\!\!\!\! \langle v^*(\eta,\mbf{k}) v(\eta,\mbf{k})\rangle = \\
&& \qquad = \int \mathcal{D} v_{\scriptscriptstyle +} \mathcal{D}v_{\scriptscriptstyle -} \, v_{\scriptscriptstyle +}^*(\eta,\mbf{k}) v_{\scriptscriptstyle -}(\eta,\mbf{k})  \exp \l( i  S_\mathrm{free}[v_{\scriptscriptstyle +},v_{\scriptscriptstyle -}] \r)  \exp \l(i S_\mathrm{int}[v_{\scriptscriptstyle +}] \r) \exp \l(-i S_\mathrm{int}[v_{\scriptscriptstyle -}] \r) \ ,\nonumber 
\ea
where $S_\mathrm{free}$ is quadratic in the fields, and $S_\mathrm{int}$ contains not just the remaining cubic and higher order terms in the action, but also any other quadratic terms we may decide to regard as perturbations. Note that there are two copies of the integration fields $v_{\scriptscriptstyle -}$ and $v_{\scriptscriptstyle +}$, because we are calculating expectation values, rather than $in$-$out$ matrix elements. This path integral expression is very useful to perturbatively expand the expectation value in powers of any interaction. In particular, each contribution can be represented by a Feynman diagram, with vertices drawn from the terms in $S_\mathrm{int}$ and propagators determined by the free action $S_\mathrm{free}$.  In our case, the latter are given by
\begin{subequations}
\ba
\parbox{20mm}{
\begin{fmfgraph*}(65,30) 
\fmfleft{i1}
\fmfright{o1}
\fmf{phantom}{i1,o1}
\fmffreeze
\fmf{vanilla}{i1,o1}
\fmfv{decor.shape=circle, decor.filled=full, decor.size=1.5thick, label=$\eta$, label.dist=9, label.angle=-90}{i1}
\fmfv{decor.shape=circle, decor.filled=full, decor.size=1.5thick, label=$\eta'$, label.angle=-90}{o1}
\end{fmfgraph*}} \quad &=& \int \mathcal{D}v_{\scriptscriptstyle +} \mathcal{D}v_{\scriptscriptstyle -} \, v^*_{\scriptscriptstyle +}(\eta,\mbf{k}) v_{\scriptscriptstyle +}(\eta',\mbf{k}) \exp(i S_\mathrm{free}) \approx \f{e^{-i k |\eta-\eta^{\pr}|}}{2k}, \label{eq:++}\\
\parbox{20mm}{
\begin{fmfgraph*}(65,30) 
\fmfleft{i1}
\fmfright{o1}
\fmf{phantom}{i1,v,o1}
\fmffreeze
\fmf{dashes}{i1,v,o1}
\fmfv{decor.shape=circle, decor.filled=full, decor.size=1.5thick, label=$\eta$, label.dist=9, label.angle=-90}{i1}
\fmfv{decor.shape=circle, decor.filled=full, decor.size=1.5thick, label=$\eta'$, label.angle=-90}{o1}
\end{fmfgraph*}} \quad  &=& \int \mathcal{D}v_{\scriptscriptstyle +} \mathcal{D}v_{\scriptscriptstyle -} \, v^*_{\scriptscriptstyle -}(\eta,\mbf{k}) v_{\scriptscriptstyle -}(\eta',\mbf{k}) \exp(i S_\mathrm{free}) \approx \f{e^{i k |\eta - \eta^{\pr}|}}{2k},   \label{eq:--} \\
\parbox{20mm}{
\begin{fmfgraph*}(65,30) 
\fmfleft{i1}
\fmfright{o1}
\fmf{phantom}{i1,v,o1}
\fmffreeze
\fmf{vanilla}{i1,v}
\fmf{dashes}{v,o1}
\fmfv{decor.shape=circle, decor.filled=full, decor.size=1.5thick, label=$\eta$, label.dist=9, label.angle=-90}{i1}
\fmfv{decor.shape=circle, decor.filled=full, decor.size=1.5thick, label=$\eta'$, label.angle=-90}{o1}
\end{fmfgraph*}} \quad &=& \int \mathcal{D}v_{\scriptscriptstyle +} \mathcal{D}v_{\scriptscriptstyle -} \, v^*_{\scriptscriptstyle +}(\eta,\mbf{k}) v_{\scriptscriptstyle -}(\eta',\mbf{k}) \exp(i S_\mathrm{free}) \approx  \f{e^{i k (\eta - \eta^{\pr})}}{2k},
\label{eq:+-} 
\ea
\end{subequations}
which we quote here just in the sub-horizon limit. Note that to first order in $S_\mathrm{int}$ there are  two vertices, one that contains powers of $v_+$ and one that contains powers of $v_-$;  the associated coefficients  just differ by an overall sign.\footnote{The quadratic action $S_\mathrm{free}$ enforces $v_+(\vec{k})=v_-(\vec{k})$ at time $\eta$.  Hence, we could replace $v_{\scriptscriptstyle +}^*(\eta,\mbf{k}) v_{\scriptscriptstyle -}(\eta,\mbf{k})$ by $v_{\scriptscriptstyle +}^*(\eta,\mbf{k}) v_{\scriptscriptstyle +}(\eta,\mbf{k})$ or $v_{\scriptscriptstyle -}^*(\eta,\mbf{k}) v_{\scriptscriptstyle -}(\eta,\mbf{k})$   inside the  path integral (\ref{eq:path integral}). Our choice removes the apparently ill-defined corrections we otherwise obtain  when higher-order time derivatives act on the time-ordered products in equations (\ref{eq:++}) and (\ref{eq:--}). These ill-defined corrections can also be eliminated by field-redefinitions, a procedure that leads to  the same corrections we find using our choice of field insertions.} 

As a simple example, let us calculate the value of the two-point function in  the short-wavelength limit to zeroth order in the interactions. Using the definition (\ref{eq:path integral}) and equation (\ref{eq:+-}),  we find
\begin{equation} \label{eq:lowest order}
	\langle v^*(\eta,\mbf{k}) v(\eta,\mbf{k})\rangle= \,\,
	\parbox{20mm}{
\begin{fmfgraph*}(65,30) 
\fmfleft{i1}
\fmfright{o1}
\fmf{phantom}{i1,v,o1}
\fmffreeze
\fmf{vanilla}{i1,v}
\fmf{dashes}{v,o1}
\fmfv{decor.shape=circle, decor.filled=full, decor.size=1.5thick, label=$\eta$, label.angle=-90}{i1}
\fmfv{decor.shape=circle, decor.filled=full, decor.size=1.5thick, label=$\eta$, label.angle=-90}{o1}
\end{fmfgraph*}} \quad
	\approx\frac{1}{2k} \quad \quad (|k\eta|\gg 1),
\end{equation}
which is the well-known and standard short-wavelength limit result. In this regime, the two-point function is hence the inverse of the dispersion relation, since the latter determines the appropriate boundary conditions for the mode functions, as in equation (\ref{20}).

In the next two sections we use the path integral (\ref{eq:path integral})  to calculate the corrections to the two-point function  coming from higher-order operators in the action. These can be interpreted as corrections to the dispersion relation, even though in the presence of such  terms  the mode equations generally contain higher order time derivatives. In any case, a significant disagreement between the calculated two-point function and  the lowest order result (\ref{eq:lowest order}) points to the  lack of self-consistency of our quantization procedure, and signals the breakdown of cosmological perturbation theory.

\section{The Limits of Perturbation Theory: Tensors}
\label{sec:tensors}

The lowest order action~(\ref{1}) contains the leading terms that describe the dynamics of the inflaton and its perturbations. However, as we have noted, in an effective field theory approach the action generically contains all possible terms compatible with general covariance and any other symmetry of the theory. Here, for simplicity, we assume invariance under parity, an approximate shift symmetry of the inflaton, and a discrete $\mathbb{Z}_2$ symmetry $\varphi\to -\varphi$. Thus, all possible effective corrections to the action (\ref{1}) can be built from the metric $g_{\mu\nu}$, the Riemann tensor $R_{\mu \nu \lambda \rho}$, the covariant derivative $\nabla_\mu$ and an even number of scalar fields $\vphi$. In what follows, we consider these additional terms and compute the corrections they induce on the two-point function of tensor perturbations in the short-wavelength limit. This allows us to determine the regime in which additional terms in the action cannot be neglected, and hence, the range over which cosmological perturbation theory is applicable. The reader not interested in technical details  may skip directly to section \ref{sec:breakdown}, where we collect and summarize our results.

\subsection{Dimension Four Operators}

We begin our analysis by considering all dimension four operators, which appear in the action multiplied by dimensionless coefficients. On dimensional grounds, we expect these to yield corrections  to the two-point function that are suppressed by only two powers\footnote{Dimension six operators quadratic in $\vphi$ also contribute at this order; we consider them later.} of $\mpl$. These operators will also help us to illustrate our formalism and discuss some of the important issues related to our calculation.

Any generally covariant  dimension four effective correction must be of the form
\be 
\la{17}
S_1 \equiv S_\alpha+S_\beta= \int \sqrt{-g} \l( \alpha R^2 + \beta C^2\r),
\ee
where $C^2$ is the square of the Weyl tensor,
\be
C^2 =  R_{\mu \nu \lambda \rho} R^{\mu \nu \lambda \rho} - 2 R_{\mu \nu} R^{\mu \nu} + \frac{1}{3}R^2,
\ee
and the dimensionless couplings $\alpha$ and $\beta$ are assumed to be of order one. Note that we have ignored total derivatives like the Gauss-Bonnet term, since they do not lead to any corrections in perturbation theory. The Levi-Civita tensor cannot appear in the action because we assume invariance under parity.

We start by substituting the perturbed metric (\ref{3}) into equation (\ref{17}) and expanding up to second order in $h_{i j}$.  Using the modified background equations and equation~(\ref{11}) to express the tensor perturbations in terms of the  variable $v$, we obtain in the sub-horizon limit
\bea
\delta_2 S_{\alpha} &=& \f{\alpha}{2 \mpl^2} \sum_{\mathbf{k}} \int d\eta' \l\{ - \f{6 a^{\pr \pr}}{a^3} v_k  \l[ v^{\pr \pr}_{-k} +  k^2  v_{-k} \r] -\f{6 a^{\pr \pr}}{a^3} \l[ v^{\pr \pr}_k + k^2 v_k\r] v_{-k}  \r\}  \la{18} , \\
\delta_2 S_{\beta} &=&  \f{\beta}{\mpl^2} \sum_{\mathbf{k}} \int d\eta' \l\{ \f{1}{a} \l[ v^{\pr \pr}_k +k^2 v_k\r] - 2 \, a H \l( \f{v_k}{a}\r)^{\pr} \r\} \cdot \l\{ \f{1}{a} \l[ v^{\pr \pr}_{-k} + k^2 v_{-k} \r] - 2 \, a H \l( \f{v_{-k}}{a}\r)^{\pr} \r\}. \nonumber
\eea
From these expressions, it is easy to derive the rules for the vertices 
\begin{subequations}
\ba
\parbox{20mm}{
\begin{fmfgraph*}(45,30)
\fmfleft{i1}
\fmfright{o1}
\fmf{vanilla}{i1,v,o1}
\fmfv{decor.shape=cross, decor.filled=full, decor.size=4thick, label=$\alpha$, label.angle=90}{v}
\end{fmfgraph*}}  \!\!\!\!\!\!  &\approx& \f{i \alpha}{\mpl^2} \int_{- \infty}^{\eta} d \eta^{\prime}  \l\{ - \f{6 a^{\pr \pr}}{a^3}  \l(\overrightarrow{\d}_{\eta^{\pr}}^2 + k^2 \r) - \l(\overleftarrow{\d}_{\eta^{\pr}}^2 + k^2 \r) \f{6 a^{\pr \pr}}{a^3} \r\}  \nonumber \\
\parbox{20mm}{
\begin{fmfgraph*}(45,30)
\fmfleft{i1}
\fmfright{o1}
\fmf{dashes}{i1,v,o1}
\fmfv{decor.shape=cross, decor.filled=full, decor.size=4thick,label=$\alpha$,label.angle=90}{v}
\end{fmfgraph*}}  \!\!\!\!\!\!  &=& - \!\!  \quad 
\parbox{20mm}{
\begin{fmfgraph*}(45,30)
\fmfkeep{avertex}
\fmfleft{i1}
\fmfright{o1}
\fmf{vanilla}{i1,v,o1}
\fmfv{decor.shape=cross, decor.filled=full, decor.size=4thick, label=$\alpha$, label.angle=90}{v}
\end{fmfgraph*}}   \\
\parbox{20mm}{
\begin{fmfgraph*}(45,30)
\fmfleft{i1}
\fmfright{o1}
\fmf{vanilla}{i1,v,o1}
\fmfv{decor.shape=cross, decor.filled=full, decor.size=4thick,label=$\beta$,label.angle=90}{v}
\end{fmfgraph*}}  \!\!\!\!\!\!  &\approx& \f{2 i \beta}{\mpl^2} \int_{- \infty}^{\eta} d \eta^{\pr} \l\{ \l(\overleftarrow{\d}_{\eta^{\pr}}^2 + k^2 \r) - \overleftarrow{\d}_{\eta^{\pr}} 2 \, a H \r\}   \f{1}{a^2} \l\{ \l( \overrightarrow{\d}_{\eta^{\pr}}^2 + k^2 \r) - 2 \, a H\overrightarrow{\d}_{\eta^{\pr}} \r\}   \nonumber \\
\parbox{20mm}{
\begin{fmfgraph*}(45,30)
\fmfleft{i1}
\fmfright{o1}
\fmf{dashes}{i1,v,o1}
\fmfv{decor.shape=cross, decor.filled=full, decor.size=4thick,label=$\beta$,label.angle=90}{v}
\end{fmfgraph*}}  \!\!\!\!\!\!  &=& - \!\!  \quad 
\parbox{20mm}{
\begin{fmfgraph*}(45,30)
\fmfkeep{avertex}
\fmfleft{i1}
\fmfright{o1}
\fmf{vanilla}{i1,v,o1}
\fmfv{decor.shape=cross, decor.filled=full, decor.size=4thick, label=$\beta$, label.angle=90}{v}
\end{fmfgraph*}}, 
\ea
\end{subequations}
where the arrows indicate the propagator on which the derivatives act (because the vertex is quadratic, two propagators meet at the vertex.)

We are now ready  to  consider the correction  due to the square of the Ricci scalar. The first order correction to the two-point function is given by the sum of the following two graphs,
\ba
\parbox{20mm}{
\begin{fmfgraph*}(80,30)
\fmfleft{i1}
\fmfright{o1}
\fmf{vanilla}{i1,v1}
\fmf{vanilla}{v1,v}
\fmf{vanilla}{v,v2}
\fmf{dashes}{v2,o1}
\fmfv{decor.shape=cross, decor.filled=full, decor.size=4thick, label=$\alpha$, label.angle=90}{v}
\fmfv{decor.shape=circle, decor.filled=full, decor.size=1.5thick, label=$\eta$, label.angle=-90}{i1,o1}
\end{fmfgraph*}} \qquad\,\, &\approx& \f{i \alpha}{\mpl^2}  \int_{- \infty}^{\eta} d \eta^{\prime}  \l\{ \f{i \delta(\eta - \eta')}{2k} \f{6 a^{\pr \pr}}{a^3} \r\} \approx - \f{\alpha}{2k}  \f{12 H^2}{\mpl^2}, 
\nonumber  \\
\parbox{20mm}{
\begin{fmfgraph*}(80,30)
\fmfleft{i1}
\fmfright{o1}
\fmf{vanilla}{i1,v1}
\fmf{dashes}{v1,v}
\fmf{dashes}{v,v2}
\fmf{dashes}{v2,o1}
\fmfv{decor.shape=cross, decor.filled=full, decor.size=4thick, label=$\alpha$, label.angle=90}{v}
\fmfv{decor.shape=circle, decor.filled=full, decor.size=1.5thick, label=$\eta$, label.angle=-90}{i1,o1}
\end{fmfgraph*}} \qquad\,\, &=&  \l( \,\,
\parbox{20mm}{
\begin{fmfgraph*}(80,30)
\fmfleft{i1}
\fmfright{o1}
\fmf{vanilla}{i1,v1}
\fmf{vanilla}{v1,v}
\fmf{vanilla}{v,v2}
\fmf{dashes}{v2,o1}
\fmfv{decor.shape=cross, decor.filled=full, decor.size=4thick, label=$\alpha$, label.angle=90}{v}
\fmfv{decor.shape=circle, decor.filled=full, decor.size=1.5thick, label=$\eta$, label.angle=-90}{i1,o1}
\end{fmfgraph*}} \qquad\,\,\r)^{\!\! *}  , \la{22}
\ea
where we have used the fact that $a'' / a^3 \approx 2 H^2$ to lowest order in slow-roll. Notice that the operator $\left( \overrightarrow{\d}_{\eta^{\pr}}^2 + k^2 \right)$ acting on the time-ordered propagators (\ref{eq:++}) or (\ref{eq:--}) produces a delta function, since both  are Green's functions.  On the other hand, when the same operator acts on the propagator (\ref{eq:+-}) we get zero, because  the latter is a regular solution of the free equation of motion (\ref{30}) in the sub-horizon limit. This remark will turn out to be very useful when studying higher dimension operators.

We can now consider the correction due to the square of the Weyl tensor. In this case,  the first order contribution is given by the sum of the following two graphs
\begin{eqnarray}
\parbox{20mm}{
\begin{fmfgraph*}(80,30)
\fmfleft{i1}
\fmfright{o1}
\fmf{vanilla}{i1,v1}
\fmf{vanilla}{v1,v}
\fmf{vanilla}{v,v2}
\fmf{dashes}{v2,o1}
\fmfv{decor.shape=cross, decor.filled=full, decor.size=4thick, label=$\beta$, label.angle=90}{v}
\fmfv{decor.shape=circle, decor.filled=full, decor.size=1.5thick, label=$\eta$, label.angle=-90}{i1,o1}
\end{fmfgraph*}} \qquad\,\, &=& -\f{2 i \beta}{\mpl^2} \int\limits_{- \infty}^{\eta} d \eta^{\prime}  \l\{  \delta(\eta - \eta') a H  + H^2  e^{2 i k (\eta' - \eta)} \r\} 
\approx - \f{\beta}{2k} \l\{ \f{4 i H k_{ph}}{\mpl^2} + \f{2 H^2}{\mpl^2}\r\} \la{35} \nonumber \\
\parbox{20mm}{
\begin{fmfgraph*}(80,30)
\fmfleft{i1}
\fmfright{o1}
\fmf{vanilla}{i1,v1}
\fmf{dashes}{v1,v}
\fmf{dashes}{v,v2}
\fmf{dashes}{v2,o1}
\fmfv{decor.shape=cross, decor.filled=full, decor.size=4thick, label=$\beta$, label.angle=90}{v}
\fmfv{decor.shape=circle, decor.filled=full, decor.size=1.5thick, label=$\eta$, label.angle=-90}{i1,o1}
\end{fmfgraph*}} \qquad\,\, &=&  \l( \,\,
\parbox{20mm}{
\begin{fmfgraph*}(80,30)
\fmfleft{i1}
\fmfright{o1}
\fmf{vanilla}{i1,v1}
\fmf{vanilla}{v1,v}
\fmf{vanilla}{v,v2}
\fmf{dashes}{v2,o1}
\fmfv{decor.shape=cross, decor.filled=full, decor.size=4thick, label=$\beta$, label.angle=90}{v}
\fmfv{decor.shape=circle, decor.filled=full, decor.size=1.5thick, label=$\eta$, label.angle=-90}{i1,o1}
\end{fmfgraph*}} \qquad\,\,\r)^{\!\! *}  . 
\end{eqnarray}
Note that the imaginary parts cancel once we sum the two graphs. This result is quite general and ensures that only corrections with even powers of $k_{ph}$ appear. 

In conclusion, we have found that the leading corrections due to dimension four operators result in a two-point function which in the short-wavelength limit has the form
\be
\parbox{20mm}{
\begin{fmfgraph*}(65,30) 
\fmfleft{i1}
\fmfright{o1}
\fmf{phantom}{i1,v,o1}
\fmffreeze
\fmf{vanilla}{i1,v}
\fmf{dashes}{v,o1}
\fmfv{decor.shape=circle, decor.filled=full, decor.size=1.5thick, label=$\eta$, label.angle=-90}{i1}
\fmfv{decor.shape=circle, decor.filled=full, decor.size=1.5thick, label=$\eta$, label.angle=-90}{o1}
\end{fmfgraph*}}  \quad + \l( \,\,
\parbox{20mm}{
\begin{fmfgraph*}(65,30)
\fmfleft{i1}
\fmfright{o1}
\fmf{vanilla}{i1,v1}
\fmf{vanilla}{v1,v}
\fmf{vanilla}{v,v2}
\fmf{dashes}{v2,o1}
\fmfv{decor.shape=cross, decor.filled=full, decor.size=4thick, label=$\alpha$, label.angle=90}{v}
\fmfv{decor.shape=circle, decor.filled=full, decor.size=1.5thick, label=$\eta$, label.angle=-90}{i1,o1}
\end{fmfgraph*}} \quad + \,\,
\parbox{20mm}{
\begin{fmfgraph*}(65,30)
\fmfleft{i1}
\fmfright{o1}
\fmf{vanilla}{i1,v1}
\fmf{vanilla}{v1,v}
\fmf{vanilla}{v,v2}
\fmf{dashes}{v2,o1}
\fmfv{decor.shape=cross, decor.filled=full, decor.size=4thick, label=$\beta$, label.angle=90}{v}
\fmfv{decor.shape=circle, decor.filled=full, decor.size=1.5thick, label=$\eta$, label.angle=-90}{i1,o1}
\end{fmfgraph*}} \quad + \mbox{c.c.} \r)
\approx \f{1}{2k} \l[ 1 - (24 \alpha +4 \beta) \, \f{H^2}{\mpl^2} \r]. 
\ee
Thus,  when $H$ becomes of order $\mpl$, these corrections  become as important as the leading result, and standard cosmological perturbation theory ceases to be applicable, as the reader may have expected.  

\subsection{Higher Dimension Operators}

We would now like to consider a generic operator of dimension $2d+4$, suppressed by a factor of order $1/\mpl^{2d}$. However, it turns out that considering directly  corrections to the action (\ref{5}) for the perturbations is a much more efficient approach than starting from generally covariant effective terms added to the Lagrangian (\ref{1}), particularly if we are interested in  identifying the dominant corrections in the sub-horizon limit. Hence, we shall focus directly on modifications to the action for the perturbations.   A related approach has been described in \cite{Cheung:2007st}.

Dimensional analysis implies that  any operator of dimension $2d+4$, quadratic in the dimensionless tensor perturbations $h_{ij}$ and proportional to $2 f$ powers of the inflaton field $\varphi$ must contain $2d-2f+4$ derivatives $\d_{\mu}$ acting on $h_{ij}$, $\varphi_0(\eta)$ and $a(\eta)$. The derivatives can be distributed and contracted using the Minkowski metric in many different ways,\footnote{The reader  may think that derivatives could be contracted not only among each other with the Minkowski metric, but also by using the additional tensor structure provided by the metric perturbations $\delta g_{\mu \nu}/a^2 = h_{i j} \eta_{\mu i} \eta_{\nu j}$. However, it turns out that $\l( \delta g_{\mu \nu} / a^2 \r) \d^{\nu} a = h_{i j} \eta_{\mu i} \d^j a = 0$ and, since $h_{i j}$ is transverse, $$ \d^{\mu} \l( \delta g_{\mu \nu} / a^2 \r) = \eta_{\nu j} \d^i h_{i j} = 0 \ .$$ Thus, we get a non-vanishing contribution only when we contract derivatives  with the Minkowski metric while the factors $\eta_{\mu i} \eta_{\nu j}$ are contracted among each other yielding an irrelevant overall factor.} but each of these terms can be schematically represented as 
\be \la{95}
\mpl^{-2 d-2} \left( \partial^{\, 2n + m + p} \, [a,\vphi_0] \right) \, \left(\partial^{\, 2q + m + r} \, v \right) \, \left(\partial^{\, 2s + p + r} \, v \ \right) ,
\ee
where we have used equation (\ref{11}) and $\partial^i[a,\vphi_0]$ is just a symbol that represents any combination of $i$ derivatives acting on  $a$'s and $\vphi_0$'s. One such term would have $2n + m + p$ derivatives acting on one or more factors of $a$ or $\vphi_0$,  $2q + m + r$ derivatives acting on one field $v$ and $2s + p + r$ acting on the other $v$. In particular, $2n$ of the derivatives acting on the scale factor or the background field are contracted among themselves while $m$ and $p$ of them are contracted with derivatives acting on, respectively, the first and second field $v$. The derivatives acting on the fields $v$ are organized in a similar way. 

Let us illustrate this notation by considering  a term with $p = s = f= 0$, $m = n = q= 1$ and $r =2$. Dimensional analysis implies that $d=3$, and thus the explicit form of such a term would be
\be
\mpl^{-8} \partial_{\, 2 + 1 + 0} \, [a] \, \partial_{\, 2 + 1 + 2} \, v \, \partial _{\, 0 + 0 + 2} \, v = \mpl^{-8} \d_{\mu} \d^{\mu} \, \d_{\nu} [a] \, \d_{\lambda} \d^{\lambda} \, \d^{\nu} \, \d_{\alpha} \d_{\beta} \, v \, \d^{\alpha} \d^{\beta} \, v \ ,
\ee
where $\d_{\mu} \d^{\mu} \d_{\nu} [a] $ denotes all possible ways to construct a term with three derivatives of the scale factor, with the given tensor structure.

The first step to estimate the leading correction due to a term of the form (\ref{95}) is realizing that this can always be expressed as a linear combination of terms of the form
\be \la{1010}
\mpl^{-2d-2} \d^{2j + l} \l[ a, \vphi_0 \r] \d^{2m + l} v \, \d_{\mu} \d^{\mu} v,
\ee
plus, possibly, a term with no derivatives acting on $v$, which in any case gives a contribution that is always subdominant in the sub-horizon limit. For a proof that this decomposition is always possible, we refer the reader to Appendix \ref{app}. In what follows, we therefore restrict ourselves to  terms of the form (\ref{1010}). 

Dimensional analysis requires that the indexes $j$, $l$ and $m$ in equation (\ref{1010}) obey
\be \la{1011}
j + l + m = d - f + 1 \ .
\ee
Furthermore, since $d^n \varphi_0/d \eta^n \propto \sqrt{2\epsilon}\, \mpl \, a^n H^n$ and $d^n a/d \eta^n \propto a^{n+1} H^n$,
each field $\vphi_0$ yields a factor of $\mpl$, while each derivative acting on it or on the scale factor results in a factor of $H$ to leading order in slow-roll. Finally, the $l$ partial derivatives $\d _{\mu}$ acting on $v$ that are contracted with derivatives acting on $a$ or $\vphi_0$ can be turned into derivatives with respect to $\eta$ only. Thus, (\ref{1010}) can be re-written as
\be
\mpl^{-2d +2f -2} f(a) H^{2j + l} \, \square^m \d_{\eta}^{\, l}  v \, \square \, v \ ,
\ee
where we have defined $\square \equiv \d_{\mu} \d^{\mu}$; the corresponding correction to the two-point function is schematically given by
\ba
\parbox{20mm}{
\begin{fmfgraph*}(70,30)
\fmfleft{i1}
\fmfright{o1}
\fmf{vanilla}{i1,v1}
\fmf{vanilla}{v1,v}
\fmf{vanilla}{v,v2}
\fmf{dashes}{v2,o1}
\fmfv{decor.shape=cross, decor.filled=full, decor.size=4thick, label.angle=90}{v}
\fmfv{decor.shape=circle, decor.filled=full, decor.size=1.5thick, label=$\eta$, label.angle=-90}{i1,o1}
\end{fmfgraph*}}  \qquad\,\, &=& \f{i}{\mpl^{2d-2f+2}} \int_{- \infty}^{\eta} d\eta' f(a) H^{2j+l} \times  \\
&& \qquad \qquad \times \,\,
\parbox{20mm}{
\begin{fmfgraph*}(65,30) 
\fmfleft{i1}
\fmfright{o1}
\fmf{phantom}{i1,o1}
\fmffreeze
\fmf{vanilla}{i1,o1}
\fmfv{decor.shape=circle, decor.filled=full, decor.size=1.5thick, label=$\eta$, label.dist=9, label.angle=-90}{i1}
\fmfv{decor.shape=circle, decor.filled=full, decor.size=1.5thick, label=$\eta'$, label.angle=-90}{o1}
\end{fmfgraph*}} \quad\quad\,
\l( \overleftarrow{\square}^m \overleftarrow{\d}_{\eta'}^{\, l} \overrightarrow{\square} + \overleftarrow{\square} \overrightarrow{\d}_{\eta'}^{\, l} \overrightarrow{\square}^m \r) \,
\parbox{20mm}{
\begin{fmfgraph*}(65,30) 
\fmfleft{i1}
\fmfright{o1}
\fmf{phantom}{i1,v,o1}
\fmffreeze
\fmf{vanilla}{i1,v}
\fmf{dashes}{v,o1}
\fmfv{decor.shape=circle, decor.filled=full, decor.size=1.5thick, label=$\eta'$, label.angle=-90}{i1}
\fmfv{decor.shape=circle, decor.filled=full, decor.size=1.5thick, label.dist=9, label=$\eta$, label.angle=-90}{o1}
\end{fmfgraph*}}\nonumber 
\ea
plus the complex conjugate of this graph. Because (\ref{eq:+-}) satisfies the free equation of motion, this correction is non--vanishing only when the index $m$ is equal to zero, and in this case we obtain
\ba
\parbox{20mm}{
\begin{fmfgraph*}(70,30)
\fmfleft{i1}
\fmfright{o1}
\fmf{vanilla}{i1,v1}
\fmf{vanilla}{v1,v}
\fmf{vanilla}{v,v2}
\fmf{dashes}{v2,o1}
\fmfv{decor.shape=cross, decor.filled=full, decor.size=4thick, label.angle=90}{v}
\fmfv{decor.shape=circle, decor.filled=full, decor.size=1.5thick, label=$\eta$, label.angle=-90}{i1,o1}
\end{fmfgraph*}}  \quad \,\,=  \f{1}{\mpl^{2d-2f+2}} \int\limits_{- \infty}^{\eta} d\eta' f(a) H^{2j+l} \delta (\eta - \eta') \f{(i k)^l}{2k} = \f{f(a)}{2k} \f{H^{2j+l} (i k )^l}{\mpl^{2d -2f +2}} . 
\ea

The leading correction in the short-wavelength limit is the one with the maximum number of powers of $k$. According to equation (\ref{1011}), this maximum number simply equals $d~-~f~+~1~\equiv~l_{max}$, and it corresponds to the case in which $j=m=0$. Thus, if $d-f$ is odd, $l_{max}$ is even and the leading correction is simply given by
\be \la{lead1}
\delta \langle v^*(\mbf{k}) v(\mbf{k})\rangle \propto \frac{1}{2k} \, \left( \frac{H}{
\mpl} \right)^{d-f+1} \, \left(\f {k_{ph}}{\mpl}\right)^{d-f+1} \qquad \qquad ( d -f \mbox{ odd}) ,
\ee
since each factor of $k/\mpl$ must be accompanied by a factor of $a$ to render the spatial momentum physical. On the other hand, if $d -f$ is even, $l_{max}$ as defined above is odd and the term with the highest number of powers of $k$ is purely imaginary. As we have seen in the previous section, such a term disappears when we add the contribution from the complex conjugate graph. Therefore, the leading correction corresponds to the largest {\it even} value of $l$, which turns out to be $l_{max} = d-f$, and is therefore given by
\be \la{lead2}
\delta \langle v^*(\mbf{k}) v(\mbf{k})\rangle \propto\frac{1}{2k}\, \left(\frac{H}{\mpl} \right)^{d-f+2} \, \left( \f {k_{ph}}{\mpl} \right)^{d-f} \qquad \qquad \,\,\, \,\,( d -f \mbox{ even})\ .
\ee

Equations (\ref{lead1}) and (\ref{lead2}) represent the main results of this section: they express the leading corrections to the two-point function (in the sub-horizon limit) associated with a generic operator of dimension $2d +4$ containing $2f$ powers of the inflaton field.
Since we have assumed an approximate shift invariance of the inflaton, the total number of derivatives, $2d-2f+2$, must be greater or equal than the number of fields $2f$, which in turn implies that $d \geq f$. Thus, we can label all the possible corrections with the non-negative index $d-f$. Their magnitude is given in Table \ref{table:corrections} for the first eight values of $d-f$.  Note that corrections with $d-f=0$ arise from the operators identified by Weinberg in \cite{Weinberg:2008hq}. The leading momentum-dependent corrections are given by operators with $d-f=1$.
\begin{table}
\begin{center}
\begin{tabular}{c|c|c|c|}
\: $d-f$ \: & \: Leading correction \: & \: $d-f$ \: & \: Leading correction \: \\
\hline
$0$ & $H^2/\mpl^2$ & $4$ & $H^6 k_{ph}^4 /\mpl^{10}$ \\
\hline
$1$ & $H^2 k_{ph}^2/\mpl^4$ & $5$ & $H^6 k_{ph}^6/\mpl^{12}$ \\
\hline
$2$ & $H^4 k_{ph}^2 /\mpl^6$ & $6$ & $H^8 k_{ph}^6/\mpl^{14}$ \\
\hline
$3$ & $H^4 k_{ph}^4 /\mpl^8$ & $7$ & $H^{8} k_{ph}^8/\mpl^{16}$ \\
\hline
\end{tabular}
\end{center}
\caption{Leading corrections to the gravitational wave two-point functions in the short-wavelength limit. \label{table:corrections}}
\end{table}

So far, we have calculated the largest possible corrections to the two-point function in the sub-horizon limit given a certain value of $d-f$. However, the reader might still wonder whether such terms can be actually obtained from a covariant action. Employing the same technique we used to study the impact of the lowest order terms, it is indeed possible to show---after some rather lengthy calculations---that the following family of covariant terms generates this kind of contributions,
\ba 
&d-f=0:& \qquad R^{\mu \nu} \, R_{\mu\nu} 
\nonumber \\
&d-f=1:& \qquad (\nabla^{\alpha} R^{\mu \nu}) \, \nabla_{\alpha} R_{\mu \nu} \\ \label{eq:tensor family}
&d-f=2:& \qquad  (\nabla^{\alpha} \nabla^{\beta} R^{\mu \nu}) \, \nabla_{\alpha} \nabla_{\beta} R_{\mu\nu}
\nonumber \\
&\vdots& \quad \quad \quad  \quad \quad \vdots  \qquad \qquad \qquad . \qquad \qquad \qquad \qquad \,\,\, \nonumber
\ea
It can be also verified that the  $d-f=1$ term yields a correction to the two-point function  proportional to the slow-roll parameter $\epsilon$, and given the structure of this family of operators, we anticipate the  remaining terms to share the same slow-roll suppression.

\subsection{The Breakdown of Perturbation Theory}\label{sec:breakdown}

The corrections to the two-point function are functions of  two dimensionful parameters, $H$ and $k_{ph}$. For our purposes, it is convenient to organize these corrections in powers of $k_{ph}$. Thus, following  Table \ref{table:corrections}, and reintroducing the subleading terms that we previously neglected,  we find that the two-point function is
\begin{eqnarray} \label{eq:1000}
\langle v^*(\mbf{k}) v(\mbf{k})\rangle\approx
\frac{1}{2k}
\Bigg[\left(1+ \alpha_{20}\frac{H^2}{\mpl^2}+\cdots\right)+  \left(\alpha_{22}\f{H^2}{\mpl^2}+\alpha_{42}\frac{H^4}{\mpl^4}+\cdots\right) \frac{k_{ph}^2}{\mpl^2} +
\nonumber \\
{}+\left(\alpha_{44}\f{H^4}{\mpl^4} +\alpha_{64}\f{H^6}{\mpl^6}+\cdots\right)
\frac{k_{ph}^4}{\mpl^4}+ \cdots  \Bigg] .
\end{eqnarray}
The coefficient $\alpha_{2 0}$ is of order one, while all the $\alpha_{n n}$ with $n \geq 2$ are of order $\epsilon$, as the family of covariant terms (\ref{eq:tensor family}) suggests. At the end of Section \ref{sec:scalars} we provide further evidence supporting this claim.

In order for Equation (\ref{eq:1000})  to be a valid perturbative expansion,  every correction term must be much smaller than one. Because $\alpha_{2 0}$ is of order one, this implies the condition 
\begin{equation} \label{eq:sub-Planckian}
	\frac{H}{\mpl} \ll 1,
\end{equation}
which must hold for all values of $k_{ph}$. Equation (\ref{eq:1000}) then shows that if condition (\ref{eq:sub-Planckian}) is satisfied, the corrections to the two-point function remain small even for $k_{ph}\approx \mpl$. In fact, to leading order in $H/\mpl$  we can rewrite equation (\ref{eq:1000}) as 
\begin{equation} \label{eq:series}
\langle v^*(\mbf{k}) v(\mbf{k})\rangle\approx
\frac{1}{2k}
\left[1 + \alpha_{22}\frac{k_{ph}^2}{\Lambda^2}+\alpha_{44}\frac{k_{ph}^4}{\Lambda^4}+ \cdots \right] \ ,
\end{equation}
where we have introduced the effective cut-off
\begin{equation}\label{eq:lambda}
	\Lambda\approx \frac{\mpl^2}{H}.
\end{equation}

Equations (\ref{eq:series}) and (\ref{eq:lambda}) are the main result of this article. For $k_{ph}\ll \Lambda$, all the corrections  are strongly suppressed and can thus be neglected. However, at $k_{ph} \approx \Lambda$, all the corrections become of order $\epsilon$, the asymptotic series breaks down, and the effective theory ceases to be valid.

To conclude this section, let us briefly comment on the effects of terms that break the shift symmetry. Because the only difference is that these terms contain undifferentiated scalars, any such correction can be cast as a generally covariant term that respects the symmetry, multiplied by a power of the dimensionless ratio $\varphi/\mpl$. Hence, these terms introduce corrections to the two-point function of the form we have already discussed, but with coefficients $\alpha_{ij}$ that can now depend on arbitrary powers of the background field $\varphi_0$,
\begin{equation}\label{eq:phi expansion}
	\alpha_{ij}=\alpha_{ij}^{(0)}+\alpha_{ij}^{(1)}\frac{\varphi_0}{\mpl}+
	\alpha_{ij}^{(2)}\left(\frac{\varphi_0}{\mpl}\right)^2+\cdots.
\end{equation}
Therefore, in the absence of any mechanism or symmetry that keeps the coefficients $\alpha_{ij}^{(1)}, \alpha_{ij}^{(2)}, \ldots$ small (e.g. an approximate shift symmetry), such an expansion looses its validity for $\varphi_0>\mpl$, regardless of the value of $k_{ph}$. If, on the other hand,  equation (\ref{eq:phi expansion}) is a sensible expansion, and  $\alpha_{ij}^{(0)}$ is much greater than $\alpha_{ij}^{(1)}, \alpha_{ij}^{(2)}, ...$, then we can effectively assume that the shift-symmetry is exact, and perturbations theory breaks down again at $k_{ph}\approx\Lambda.$

\subsection{Loop Diagrams}
Our analysis so far has concentrated on   tree-level diagrams, though loop corrections could also  invalidate the short-wavelength expansion.   The contribution of loops is obscured by the appearance of divergent momentum integrals. At any order in the derivative expansion it is still possible to cancel these divergences by renormalizing a finite number of parameters, provided that all terms consistent with the symmetries of the theory are included in the Lagrangian \cite{Weinberg:1995mt}. In practice, this cancellation is due to  the presence of appropriate counter-terms in the Lagrangian.  For this reason, divergent integrals in loop diagrams are rather harmless. They yield corrections of the same structure as tree-level diagrams, modulo a (mild) logarithmic running of their values with scale \cite{Donoghue:1994dn}. Hence, we do not expect this type of contributions to drastically change our conclusions, though we should emphasize that this is just an expectation.

\section{The Limits of Perturbation Theory: Scalars}
\label{sec:scalars}

We now turn our attention to corrections to the two-point function of scalar perturbations. Despite some complications that are particular to the this sector,   the method developed in the previous section can be easily extended to scalars.

To this end, let us consider the action $S = S_0 + \lambda S_1$, where $S_0$ is the action (\ref{1}) describing a scalar field minimally coupled to Einstein gravity, while $S_1$ is a generic generally covariant correction suppressed by a coupling $\lambda \sim 1/\mpl^{2d}$. As we pointed out in the previous section, $S_1$ generically involves contractions of the Riemann tensor $R_{\mu \nu \lambda \rho}$ and the covariant derivative $\nabla_\mu$ as well as the scalar field $\vphi$. In order to compute the resulting first order contribution to the two-point function for $v=a\, \delta\varphi$, we insert the perturbed metric (\ref{metricpert}) and the perturbed inflaton field  into the action $S$ and expand up to second order in $\delta \vphi, \phi$ and $B$. We then express the metric perturbations in terms of $\delta \vphi$  using equations (\ref{eq:constraints}) and, finally, in terms of $v$ using the definition (\ref{38}).  Note that, even though the relations (\ref{eq:constraints}) were derived by solving the constraints associated with the unperturbed action for the perturbations $\delta_2 S_0$, corrections to these relations due to $\delta_2 S_1$ do not contribute at first order in $\lambda$.
To show this, let us expand $\phi$ and $B$ in powers of the coupling
\be
\phi = \phi_0 + \lambda \phi_1 + \mathcal{O}(\lambda^2), \qquad B=B_0 + \lambda B_1 +
\mathcal{O}(\lambda^2),
\ee
where $\phi_0$ and $B_0$ satisfy the unperturbed relations (\ref{60}) and (\ref{61}) respectively. By expanding the full quadratic action for the perturbations $\delta_2 S \, [\delta \vphi, \phi, B]$ to first order in $\lambda$ we obtain
\ba \label{eq:expansion}
\delta_2 S \, [\delta \vphi, \phi, B] \approx \delta_2 S_0\, [\delta \vphi, \phi_0, B_0] &+&  \lambda \int \left[ \fr{\delta(\delta_2 S_0)}{\delta \phi} \right]_{\lambda=0}\phi_1 +  \lambda \int  \left[ \fr{\delta(\delta_2 S_0)}{\delta B}\right]_{\lambda=0}  B_1+ \nonumber \\ 
&+& \lambda \, \delta_2 S_1 [ \delta \vphi, \phi_0, B_0 ]+\mathcal{O}(\lambda^2).
\ea
However, the second and the third term vanish because they contain the unperturbed constraint equations evaluated at $\phi_0$ and $B_0$, which by assumption are solutions to the constraints. Therefore, as long as we are interested in first order results, we can safely use the unperturbed solutions $\phi_0$ and $B_0$ given by equations (\ref{eq:constraints}).

Expanding the leading action $S_0$ to quadratic order in the perturbations, we obtain the free action (\ref{5}). The additional quadratic terms stemming from $S_1$ must be appropriately contracted expressions containing partial derivatives  of the perturbation variable $v$, the scale factor $a$ and the background field $\varphi_0$.  However, in the case of scalar perturbations, the field $v$ can arise from fluctuations of the scalar field, $\delta \vphi = v/a$, or from fluctuations of the metric,
\be
\delta g^{\mu \nu} = - \f{a \sqrt{2 \epsilon}}{\mpl} \l[ 2 v_k \delta^\mu{}_0 \delta^\nu{}_0 + \f{i k_j}{k^2} \l( v_k' - a H v_k \r) \l( \delta^\mu{}_ j \delta^\nu{}_0 + \delta^\mu{}_0 \delta^\nu{}_ j \r)\r] \equiv \f{a \sqrt{2\epsilon}\, \mathcal{V}^{\mu \nu}}{\mpl} \ .
\ee
Thus, unlike the case of tensor perturbations, $\delta g^{\mu\nu}$ provides an additional tensor structure that can be used to contract derivatives. We now show that such contractions yield terms where the derivatives acting on $v$ or $a$ are contracted with $\eta^{\mu \nu}$. This means that the argument in the previous section can be applied to scalar perturbations as well, yielding essentially the same results. We note that terms which contain only fluctuations coming directly from the scalar field do not present this problem, and can be easily written as in equation (\ref{95}). 

Let us first consider terms with only one factor of $\mathcal{V}^{\mu \nu}$. In this case, $\mathcal{V}^{\mu \nu}$ can be contracted either with $\eta_{\mu \nu}$, leading to $
\mathcal{V}^{\mu \nu} \eta_{\mu \nu} = 2 \, v$, or with two derivatives $\d_{\mu} \d_{\nu}$, resulting in
\begin{subequations}
\ba
\d_{\mu} \d_{\nu} \mathcal{V}^{\mu \nu} &=&  2  \l( \f{\d_{\mu} \d^{\mu} a}{a} v - \f{\d_{\mu}a \, \d^{\mu} a}{a^2} v + \f{\d_{\mu} a \, \d^{\mu} v}{a} \r),   \\
\d_{\mu} a \, \d_{\nu} \mathcal{V}^{\mu \nu}  &=&  \d_{\mu} a  \, \d^{\mu} v + \f{\d_{\mu} a \, \d^{\mu}a}{a} v,   \la{1001}\\
\d_{\mu} \d_{\nu} [ a,\varphi_0 ]  \, \mathcal{V}^{\mu \nu} &=&   2 \,  ( \d_{\mu} \d^{\mu}  [a,\varphi_0]  ) \, v  \ ,
\ea
\end{subequations}
where, again, the square brackets in the last line mean that the derivatives can act on {\it one or more} factors of $a$ or $\varphi_0$. Thus, terms with only one factor of $\mathcal{V}_{\mu \nu}$ do not present any problem since, as anticipated, all the derivatives are contracted with the inverse of the Minkowski metric. 

Corrections which contain two factors of $\mathcal{V}^{\mu \nu}$, and are not products of terms in (\ref{1001}), can always be recast as 
\begin{subequations}
\ba
\mathcal{V}_{\mu \nu}  \mathcal{V}^{\mu \nu} &=& 2v^2 + \f{2}{k^2} \l[  \d_{\mu} v \, \d^{\mu} v - \f{\d_{\mu} a}{a} \, \d^{\mu} ( v^2 ) + \f{\d_{\mu} a \, \d^{\mu} a}{a^2} v^2 \r], \\
\d^{\mu} \mathcal{V}_{\mu \nu} \d_{\lambda} \mathcal{V}^{\lambda \nu} &=& - \f{1}{k^2} \d_{\mu} v' \d^{\mu} v' + \f{1}{k^2} \l[ 2  \f{\d_{\mu} a \, \d^{\mu} v}{a} \d_{\nu} \d^{\nu} v - 2 \l( \f{\d_{\mu} \d_{\nu} a}{a} - \f{\d_{\mu} a \, \d_{\nu} a}{a^2} \r) v \, \d^{\mu} \d^{\nu} v - \r. \nonumber \\
&& \qquad \qquad \l. \f{\d_{\mu} a \, \d^{\mu} a}{a^2} \d_{\nu} v \, \d^{\nu} v +  \l( \f{\d_{\mu} \d^{\mu} a}{a} - \f{\d_{\mu} a \, \d^{\mu} a}{a^2} \r) \f{\d_{\nu} a}{a} \, \d^{\nu} ( v^2)  \r] .\la{1003}
\ea
\end{subequations}
All the terms inside the square brackets become negligible in the sub-horizon limit, since their contribution is suppressed by an extra factor of $1/k^2$. The only term in which some derivatives are not contracted with the Minkowski inverse metric  is the first one in equation (\ref{1003}). However, the two derivatives with respect to conformal time result in a factor of $k^2$ which is precisely canceled by the extra factor $1/k^2$, and for all practical purposes such a term is equivalent to  $\d_{\mu} v \, \d^{\mu} v$.

Therefore, we have demonstrated that terms quadratic in the scalar fluctuations can be schematically written as in equation~(\ref{95}). The remainder of the analysis then proceeds as for tensor perturbations, and effective corrections to scalar perturbations are thus also subdominant in the regime
\begin{equation}
H\ll \mpl \quad \text{and}\quad  k_{ph}\ll \Lambda\sim \frac{\mpl^2}{H} \ .
\end{equation}

Before concluding, we would like to address again whether  the operators that we have considered  can be actually obtained from generally covariant terms. In the case of scalar perturbations, it is indeed possible to show---after further rather lengthy calculations---that the following family of covariant terms generates the kind of corrections shown in Table~\ref{table:corrections},
\ba
&d-f=0:& \qquad R^{\mu \nu} \, (\nabla_{\mu} \, \vphi) \, \nabla_{\nu} \, \vphi \nonumber \\
&d-f=1:& \qquad R^{\mu \nu} \, (\nabla_{\mu} \, \vphi) \, \nabla_{\nu} \, \nabla_{\gamma} \,\nabla^{\gamma} \, \vphi \la{120}  \\
&d-f=2:& \qquad (\nabla^{\alpha} \nabla^{\beta} R^{\mu \nu}) \, (\nabla_{\alpha} \nabla_{\mu} \, \vphi) \, \nabla_{\beta} \nabla_{\nu} \, \vphi \nonumber \\
&d-f=3:&\qquad (\nabla^{\alpha} \nabla^{\beta} R^{\mu \nu}) \, (\nabla_{\alpha} \nabla_{\mu} \, \vphi) \, \nabla_{\beta} \nabla_{\nu} \, \nabla_{\gamma} \,\nabla^{\gamma} \, \vphi \nonumber \\
&\vdots& \quad \quad \quad  \quad \quad \vdots \qquad \qquad \qquad  \qquad \qquad \qquad \, \, \, . \qquad \nonumber 
\ea
In order to illustrate how this happens, let us consider for example the $d-f=1$ term. It contains, among many other terms a factor
\be 
a^2 \, R^{\mu \nu} \, \d_{\mu} \, \delta  \vphi \, \d_{\nu} \, \d_{\gamma} \, \d^{\gamma} \, \delta \vphi 
\supset \f{2 \epsilon}{a^6} \, \d^{\mu} a \, \d^{\nu} a  \, \d_{\mu} \,  v \, \d_{\nu} \, \d_{\gamma} \, \d^{\gamma} \, v \sim -  \f{2 \epsilon}{a^6} \, \d^{\mu} a \, \d^{\nu} a  \, \d_{\mu}  \, \d_{\nu} \,  v \, \square \, v + ... \, \la{end}
\ee
where, in the last step, we have neglected a subdominant contribution in the short-wavelength limit. The last term in (\ref{end}) indeed generates a correction proportional to $H^2 k_{ph}^2 / \mpl^4$ and it is suppressed by one factor of the slow-roll parameter.  It is relatively easy to verify that the corrections generated by the other members of the family (\ref{120}) have the same slow-roll suppression, which
 strongly supports the assumption we made in the context of tensor perturbations. 
 
\section{Conclusions}
\label{sec:conclusions}

The connection, through cosmological inflation, between physics on the smallest scales, described by quantum field theory, and that on the largest scales in the universe is one of the most profound aspects of modern cosmology.  However, since inflation takes place at such early epochs, and magnifies fluctuations of such small wavelengths, it is important to establish the regime of validity of the usual formalism---that of semiclassical gravity, with quantum field theory assumed valid, and coupled to the minimal Einstein-Hilbert action---at those scales.

On general grounds, we expect the canonical approach to break down at ultra-short distances, where the operators  that arise in an effective field theory treatment of the coupled metric-inflaton system become relevant. In this article we have calculated the impact of these higher-dimensional operators on the power spectrum at short wavelengths.  In this way, we have been able to determine the regime in which the properties of the perturbations deviate from what is conventionally assumed. From a purely theoretical standpoint, these considerations are important if we are to understand the limits of applicability of cosmological perturbation theory. From an observational standpoint, cosmic microwave background measurements are becoming so precise that we may hope to use them to identify the signatures of new gravitational or field theoretic physics.

Our analysis has focused on tree-level corrections to the spectrum. Because we have essentially considered all possible  generally covariant terms in the effective action,  we expect to have unveiled the form of all possible corrections that are compatible with the underlying symmetries of the theory. Loop diagrams may yield additional scale-dependent logarithmic corrections neglected in our analysis. In any case, our results  indicate that cosmological perturbation theory does not apply  all the way to infinitesimally small distances, $k_{ph}\to \infty$, and that, indeed, there is a physical spatial momentum $\Lambda$ (or a physical length $1/\Lambda$) beyond which cosmological perturbation theory breaks down.  The scale at which perturbation theory breaks down at  tree-level is 
\begin{equation}\label{eq:Lambda}
\Lambda \sim \frac{\mpl^2}{H},
\end{equation}
which, because of existing limits on the scalar to tensor power spectrum ratio \cite{Spergel:2006hy},  is at least $10^4$ times the Planck scale. 

These results have significant implications for the impact of an effective cut-off  $\Lambda$  on the primordial spectrum of primordial perturbations, which typically is at most of order $H/\Lambda$ \cite{ArmendarizPicon:2003gd}. Substituting the value of $\Lambda$ we have found, we obtain corrections of the order of $H^2/\mpl^2$, which are likely to remain unobservable \cite{Okamoto:2003wk}. This value of $\Lambda$ also solves a problem that was noticed in \cite{ArmendarizPicon:2006pd}, namely, that in the presence of a Planckian cut-off, cosmological perturbations do not tend to decay into the Bunch-Davis vacuum (or similar states). In particular, to lowest order in perturbation theory, the transition probability from an excited state into the Bunch-Davis vacuum  is significantly less than one for $\Lambda=\mpl$, but proportionally larger if $\Lambda$ is given by (\ref{eq:Lambda}).  Ultimately, a large decay probability is what justifies the choice of the Bunch-Davies vacuum as the preferred initial state for the perturbations at scales below the cut-off, since, as we have found, our theories lose their validity at momentum scales above the spatial momentum $\Lambda$.

\begin{acknowledgments}
The authors thank Robert Brandenberger, Jaume Garriga and Richard Holman for useful comments and conversations. The work of CAP and RP was supported in part by the National Science Foundation (NSF) under grant PHY-0604760. The work of MF and MT was supported in part by the NSF under grants PHY-0354990 and PHY-0653563, by funds from Syracuse University and by Research Corporation. 
\end{acknowledgments}

\appendix

\section{} \la{app}
In this appendix, we show how to integrate by parts every term of the form 
\be \la{a1}
\partial^{\, 2n + m + p} \, [a,\vphi_0] \, \partial^{\, 2q + m + r} \, v \, \partial^{\, 2s + p + r} \, v
\ee
in order to express it as a linear combination of terms like
\be \la{a2}
\d^{2j + l} \l[ a, \vphi_0 \r] \d^{2m + l} v \, \square \, v 
\ee 
plus, possibly, a term with no derivatives acting on $v$. Notice that, for notational convenience, we have defined $ \square \equiv \d_{\mu} \d^{\mu}$. Of course, if the index $q$ (or $s$) in equation (\ref{a1}) is not zero, we can easily integrate by parts $2q + m + r -2$ ($2s + p + r - 2$) times to get only terms of the form of that in equation (\ref{a2}). Therefore, in what follows we only consider terms with $q=s=0$. In this case, we can always integrate by parts an appropriate number of times to get only terms for which $m=p$. Thus, without loss of generality, we can restrict ourselves to considering terms of the form
\be \la{a3}
\partial^{\, 2n + m + p} \, [a,\vphi_0] \, \partial^{\, m+ r} \, v \, \partial^{\, p +r} \, v,  \qquad \qquad (m=p) \ .
\ee
The derivatives acting on $v$ that are contracted with derivatives acting on $a$ or $\vphi_0$ can be systematically eliminated by repeated integrations by parts:
\ba
\partial^{\, 2n + m + p} \, [a,\vphi_0] \, \partial^{\, m+ r} \, v \, \partial^{\, p +r} \, v &\sim& -  \partial^{\, 2n + m + (p-1)} \, [a,\vphi_0] \, \partial^{\, m+ r} \,  v \, \partial^{\, (p-1) +r} \,\square \,  v \\ 
&& \,\, + \f{1}{2}  \partial^{\, 2 (n+1) + (m-1) + (p-1)} \, [a,\vphi_0] \, \partial^{\, (m-1)+ (r+1)} \, v \, \partial^{\, (p-1) +(r+1)} \, v \ , \nonumber
\ea
where we have denoted equivalence up to integration by parts with $\sim$. The first term on the right hand side can be cast in the form (\ref{a2}) by integrating by parts $(p -1) + r$ times, while the second one is of the form (\ref{a3}) with $n$ and $r$ ($p$ and $m$) increased (decreased) by one. By iterating this procedure, we eventually obtain terms of the form
\be \la{a4}
\partial^{\, 2n} \, [a,\vphi_0] \, \partial^{\, r} \, v \, \partial^{\, r} \, v \ ,
\ee
where now $n$ and $r$ have changed. Again, we can integrate by parts  and obtain
\ba
\partial^{\, 2n} \, [a,\vphi_0] \, \partial^{\, r} \, v \, \partial^{\, r} \, v \sim - \partial^{\, 2n} \, [a,\vphi_0] \, \partial^{\, r-1} \, v \, \partial^{\, r-1} \, \square \, v + \f{1}{2} \partial^{\, 2(n+1)} \, [a,\vphi_0] \, \partial^{\, (r-1)} \, v \, \partial^{\, (r-1)} \, v \ .
\ea
The first term on the right hand side can be re-written as (\ref{a2}) after $r-1$ integrations by parts, while the second term has the form (\ref{a4}) with $n$ ($r$) increased (decreased) by one. Thus, by repeating this procedure we obtain many terms of the form (\ref{a2}) and we are eventually left with a term without derivatives acting on $v$. This completes our proof.

\end{fmffile}

\end{document}